\documentclass{article}
\usepackage{spconf,amsmath,graphicx}
\usepackage{booktabs}
\usepackage{xcolor}
\usepackage{hyperref}

\colorlet{todocolor}{red!70!black}

\long\def\remark#1{}

\def\glnref#1{eq. (\ref{#1})}
\def\abbref#1{fig.~\ref{#1}}

\def\absref#1{section~{\bf \ref{#1}}}
\def\tabref#1{table~(\ref{#1})}

\def\L{{\cal L}}

\title{Towards Universal Neural Vocoding with a Multi-band Excited WaveNet}
\name{Axel Roebel\thanks{This work has been funded partly by the ANR project ARS (ANR-19-
CE38-0001-01). This work was performed using HPC resources from GENCI-IDRIS (Grant 2021-AD011011177R1])}, Frederik Bous}
\address{%
  Analysis/Synthesis Team - STMS Lab. \\%
  IRCAM, CNRS, Sorbonne Université  \\%
  Paris, France%
}

\begin{document}
\ninept
\maketitle
\begin{abstract}
  This paper introduces  the Multi-Band Excited WaveNet a neural  vocoder for speaking and
  singing voices.  It aims to  advance the  state of the  art towards an  universal neural
  vocoder, which  is a  model that can  generate voice  signals from  arbitrary mel
  spectrograms extracted from  voice signals. Following the  success of the DDSP  model and
  following the development of the recently  proposed excitation vocoders we propose a
  vocoder  structure  consisting  of  multiple  specialized DNN  that  are  combined  with
  dedicated signal processing components. All components are implemented as differentiable
  operators and therefore allow joined optimization  of the model parameters. To prove the
  capacity of the model  to reproduce high quality voice signals we  evaluate the model on
  single  and   multi  speaker/singer  datasets.   We  conduct  a   subjective  evaluation
  demonstrating that the models support a  wide range of domain variations (unseen voices,
  languages, expressivity) achieving perceptive quality that  compares with a state of the
  art universal neural vocoder, however  using significantly smaller training datasets and
  significantly less parameters. We also  demonstrate remaining limits of the universality
  of neural vocoders e.g. the creation of saturated singing voices.
\end{abstract}

\vspace{-1mm}
\section{Introduction}
\label{sec:intro}

The introduction of the WaveNet \cite{Shen:18} has demonstrated that DNN can be trained to
produce high  quality speech signals when  conditioned on a mel  spectrogram.  This result
has triggered numerous research activities aiming to reduce the high computational demands
of the  original WaveNet  or to  reduce the  size of  the training  data that  is required
\cite{vandenOord:17,Prenger:19, Wang:20, Yamamoto:20,Yang:21}. Recently research focus has
been extended from single speaker models to  multi speaker models or even universal neural
vocoders  \cite{Lorenzo-Trueba:19b,Jang:21,   Jiao:21}  that  is  vocoders   that  support
arbitrary speakers, languages and expressivity.   An important motivation for constructing
a universal neural vocoder  is the simplification of the process to  create new voices for
TTS systems.   An interesting  line of  research in this  context are  models that  try to
incorporate   prior   information   about   the   speech   signal   into   the   generator
\cite{Song:19,Juvela:19,Juvela:19a,Oh:20}.   These models,  in  the  following denoted  as
excitation networks, simplify  the task of the  generator by means of  splitting the vocal
tract filter  (VTF) into a  dedicated unit.  Instead of  generating the speech  signal the
generator is then used only to produce the excitation signal.  On the other hand, only one
of these models \cite{Juvela:19a}  takes a mel spectrogram as input.   The others use more
classical vocoder  parameters like F0,  line spectral frequencies, and  an voiced/unvoiced
flag.   The  idea  to  introduce  domain  knowledge  into  the  model  seems  particularly
interesting.  It  is in  line with  the recent DDSP  \cite{Engel:19a} framework  for music
synthesis that replaces  a part of the generator  by means of a sinusoidal  model and uses
the DNN only to control the parameters  of the sinusoidal model.  The main disadvantage of
using the classical vocoder parameters for  conditioning is the fact that these parameters
are  deeply entangled.   Disentangling a  set  of heterogeneous  vocoder parameters  seems
significantly more difficult  than disentangling for example the speaker  identity from the
mel  spectrogram.  This  is due  to the  fact that  the mel  spectrogram is  a homogeneous
representation similar to images and  therefore techniques for attribute manipulation that
have proven  useful for image manipulation  (notably disentanglement) can be  applied with
only minor changes.  By consequence research on voice attribute manipulation like: Speaker
Identity Conversion  \cite{Zhang:20}, rhythm  and F0  conversion \cite{Qian:20a,Qian:20b},
Gender  Conversion \cite{Benaroya:21},  speaker normalization  \cite{Biadsy:19a} generally
starts with  a (mel) spectral  representation of the voice  signal.  In a  companion paper
that  demonstrates  high  quality  singing  voice  transpositions  over  multiple  octaves
\cite{Bous:22}, the manipulation of the mel  spectrogram and the resynthesis with a neural
vocoder has proven highly effective.

These  experiences  motivate our  research  into  extending  the  voice signals  that  are
supported by  neural vocoders. The present  paper discusses especially the  case of speech
and singing signals and will present a new neural vocoder with signficantly better support
for singing than existing models. To achieve  this goal we will introduce 2 novelties that
are the central contributions of the present research:

\begin{itemize}
\item To improve the signal quality as well as to ease the use of the vocoder in practical
  applications we  will replace the  approximate VTF  estimation from the  mel spectrogram
  proposed  in \cite{Juvela:19a}  by means  of a  small separate  model that  predicts the
  cepstral coefficients of the VTF.
\item To facilitate the generation of long and stable quasi periodic oscillations that are
  crucial for singing we  simplify the task of the excitation generator by means of
  splitting the generator  into a small DNN that predicts the F0 contour from  the input mel
  spectrogram and  a differentiable  wavetable generator  that produces  the correspoding
  excitation. The subsequent WaveNet then operates without recursion and only has the task
  to shape the given periodic pulses in accordance with the conditioning mel spectrogram.
\end{itemize}

The  rest of  the paper  is  organized as  follows. In  \absref{sec:model_components} we  will
introduce  the various  components of  the model  and put  them into  context of  existing
work.   In   \absref{sec:model_topology}   we   will   describe   the   model   topology,   in
\absref{sec:experiments} will describe the datasets and will discuss our experimental results.

\vspace{-2mm}
\section{Model components}
\label{sec:model_components}

The present  section will  discuss the structure  of the proposed  neural vocoder  that we
denote as  multi band excited  WaveNet.  We will  notably discuss relations  with existing
work.  The  fundamental idea  of the  present work  follows and  extends the  arguments in
\cite{Song:19,Juvela:19,Juvela:19a, Oh:20} that the  excitation networks with objective to
simplify the task of the WaveNet generator by means of removing the VTF from the generator
output.  Similar to \cite{Juvela:19a} we use the mel spectrogram to represent the acoustic
features.  The following section describes the proposed contributions in more details.

\vspace{-2mm}
\subsection{VTF generation}
\label{sec:vtf}

\cite{Juvela:19a} proposes to recover an approximate all-pole representation of the VTF by
means of first converting  the log amplitudes in the mel  spectrogram to linear amplitudes
and then applying the pseudo-inverse of the mel filter bank to recover an approximation of
the  linear amplitude  spectrum  from which  an  all-pole  model of  the  envelope can  be
obtained. It is well known known however that all-pole estimation from harmonic spectra is
subject to systematic errors \cite{El-Jaroudi:91}.  To counter these systematic errors the
generation of  the VTF by  means of an  auxiliary DNN seems  a preferable option.  Here we
propose to use an auxiliary DNN that  predicts a cepstral representation of the VTF.  This
prediction is cheap because it is performed frame wise and operates therefore with a small
samplerate.   We limit  the model  to predict  causal cepstral  coefficients, so  that the
resulting VTF will be minimum phase \cite{Smith:11}.

Whether we use all-polse VTF or cepstral representations, in both cases when we predict the
VTF from the Mel spectrogram we encounter the question of the gain coefficients. If we do
not constrain the VTF we create a gain ambiguity because any gain in the VTF can be
compensated by a inverse gain in the excitation generator. We therefore decide to force the
cepstral model to have zero gain.

\vspace{-2mm}
\subsection{Wavetable based excitation generation}
\label{sec:excitation}

The existing excitation networks  all use a WaveNet or more generally a  DNN to create the
quasi periodic excitation. In our experiments  notably for singing signals we noticed that
the  generation of  a stable  quasi periodic  excitation is  a difficult  problem for  the
generator.   For our  neural  vocoder we  therefore  decided to  create  a quasi  periodic
excitation  and pass  this  excitation together  with  a white  noise  signal through  the
WaveNet. Given the  F0 contour is already  correct the WaveNet now serves  only create the
appropriate  pulse  form  and  the   balance  between  the  deterministic  and  stochastic
components.

Predicing  the F0  from the  mel spectrum  has turned  out to  be rather  simple. For  the
generation of the quasiperiodic excitation we decided to use wavetable synthesis aiming to
create  an excitation  that  is approximately  white  so that  all  harmonics are  already
present.  The problem here  is that the number of harmonics depends on  the F0.  To ensure
the time varying number of harmonics we create N wavetables for $N$ different F0 ranges (n=5
for the present study). The band limited pulses for each of the $N$ wavetables are generated
in the  spectral domain depending on  the F0 range such  that even for the  maximum F0 for
that a  avtable entry will be  used no aliasing  takes place. The wavetable  synthesis are
always performed in  parallel and the final  excitation is generated by  means of linearly
interpolating only two  of the $N$ available wavetables.  The whole  process is implemented to
run on  the GPU.   The wavetable positions  $P_k$ that  need to be  sampled to  ensure the
correct F0 contour are
\begin{equation}
  P_k = N (\sum_{i=0}^{k}(F_k/R) \mathbin{\%} 1),
  \label{eqn:wavetable_inc}
\end{equation}
where the  $N$ is  the size  of the  wavetable, $F_k$  the F0  contour in  Hz and  $R$ the
samplerate in Hz.  Because $F_k$ is a continuous  variable the values to be taken from the
wavetable will not fall onto the grid of  values stored in the tables. For the gradient to
pass through the wavetable into the F0  predictor it is important that the positions $P_k$
are not quantized!  Instead the values to be output from the wavetable need to be linearly
interpolated.

\begin{figure}[htb]
  \vspace{-8mm}
  \centering
  \includegraphics[width=8.5cm]{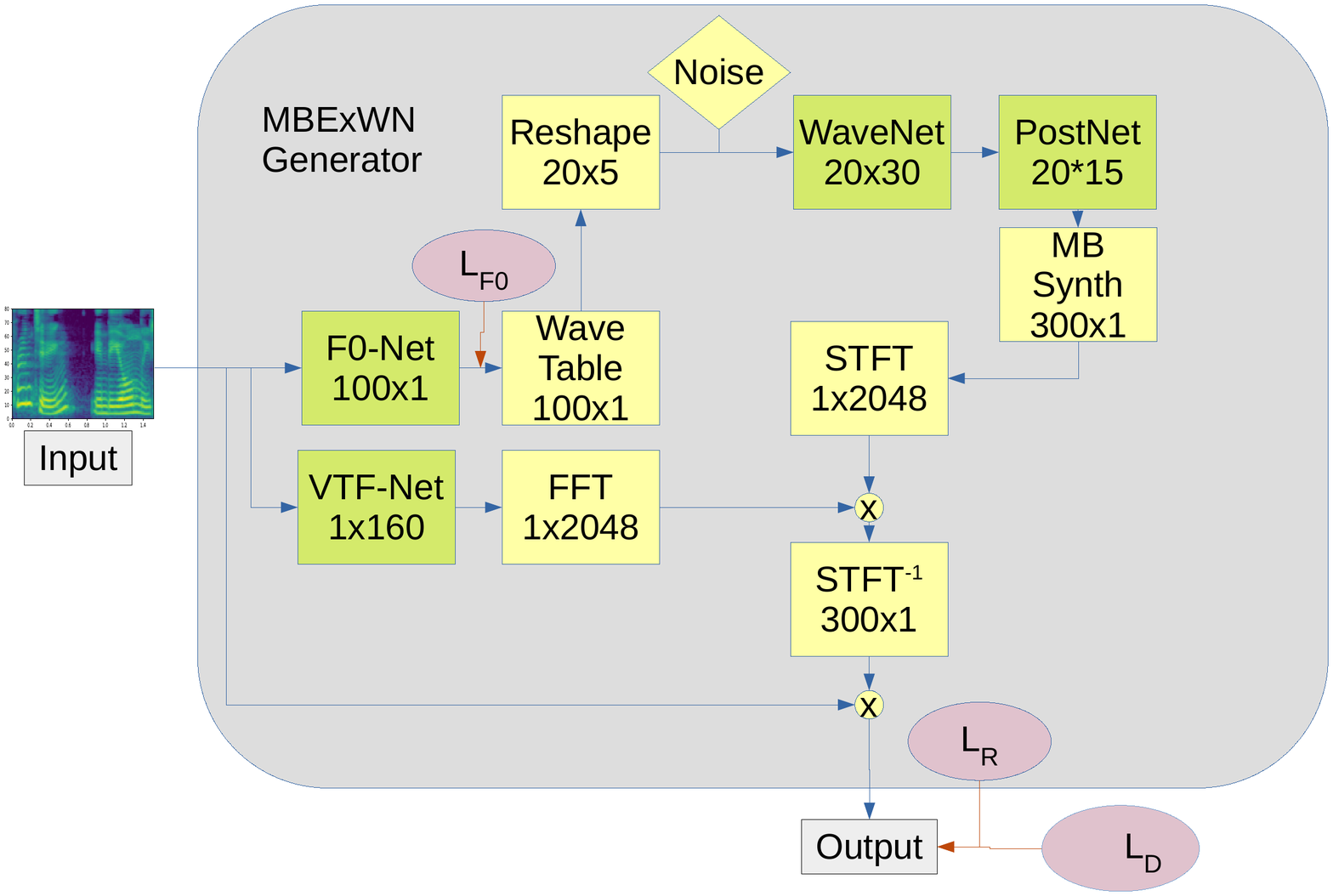}
  \vspace{-9mm}
  \caption{MBExWN  schematic generator:  Green  boxes  are DNN  models,  yellow boxes  are
    differentiable operators,  red ovals are losses.  The numbers below the  boxes specify
    the output dimensions  of the box in  the format time x channels  (batch dimension not
    shown). }
\label{fig:generator}
\end{figure}

\vspace{-4mm}
\section{Model topology}
\label{sec:model_topology}

The overall schematic diagramm of the MBExWN generator is shown in \abbref{fig:generator}.
The diagram represents the flow of a single  frame of a mel spectrogram with  80 channels. Each
block dispays the output  dimensions it would  produce for the  single spectral  frame. The
signal flow  will be discussed  in more detail  below. 

The input Mel  spectrogram enters three subnets:  First the F0 subnet that  produces an F0
sequence  with upsampling  factor 100.  The  sequence of  layers  of the  F0 predictor  is
specified  in terms  of layer  type  (C:Conv1D, L:linear  upsamling) followed  by a  short
parameter spefication. The  Conv1D layer parameters are  given as kernel size  x number of
filters optionally followed by an upsamplingfactor.   If given the upsampling performed by
means  of reshaping  channels  into time  dimension.   As an  example  consider the  layer
specification C:3x240x2.  This would be implemented by means of a Conv1D layer with kernel
size 3 and 120 channels followed b a reshape operation that upsamples by factor 2 by means
of folding every other channel into time direction.  The linear interpolation layer L is a
Conv1d layer with  precopted parameters that performes upsampling. The  only paramter here
is the upsampling factor.

The  F0  net specfication  is  then  as  follows:  C:3x150, C:3x300x2,  C:5x150,  C:3x120,
C:3x600x5, C:1x120, C:3x500x5, C:1x100, C3:50,L:2

The activation  functions in the  F0 predictor  are all relu  and are situated  after each
convolutional layer. The only excpetion here is the last layer that uses a soft sigmoid as
activation  function. The  output vector  is  then offset  and  scaled to  the desired  F0
range. In the presetn model this range is 45Hz-1400Hz.  After this operation the F0
contour passes through the wavetable generator described in \absref{sec:excitation}. It follows
a reshape operation  and a concatenation of  a white noise signal duplicating  the size of
the  excitaion  signal.   The  basic  excitation signal  then  enters  the  pulse  shaping
WaveNet.  This  WaveNet  is  following   the  classical  configuration  using  gated  tanh
activations and  kernel size 3.  It consists of  2 blocks of 5  layers, having 240  or 320
channels for single or multi voice models.

The PostNEt is a single Conv1D layer that reduces the channel size from 30 to 15 to adapt
the WaveNet output for the subsequent  a PQMF \cite{Yang:21} synthesis filter with 15
bands.

The VTF predictor is again a CNN with the specification: C:3x400, C:1x600, C:1x400,
C:1x400, C:1x160.

Activations functions are relu after all but the last convolutional layer. The final layer 
does not have any activation function and passes directly into a real valued FFT operator
to produce a minimum phase spectral envelope \cite{Smith:11}.

VTF and excitation signal produced by the PostNet are multiplied in the spectral domain
to produce the final speech signal. The STFT parameter are copied from the parameters used
for creating the mel spectrogram. 

\vspace{-2mm}
\section{Experiments}
\label{sec:experiments}

For  the following  experiments we  used 4  databases. The  first is  the LJSpeech  single
speaker dataset \cite{Ito:17} denoted as \emph{LJ}  in the following.  The second, denoted
as  \emph{SP}, is  a multi  speaker dataset  composed of  VCTK \cite{Yamagishi:19},  PTDB
\cite{Pirker:11} and AttHack \cite{LeMoine:20}  datasets.  The \emph{SP} dataset contains
approximately 45h of  speech recorded from 150 speakers.  For  singing voice experiments
we used a single singer dataset  containing a greek byzantine singer \cite{Grammalidis:16}
denoted  as \emph{DI}  and  for the  multi  singer model  a database  composed  of the  NUS
\cite{Duan:13}, SVDB \cite{Tsirulnik:19}, PJS \cite{Koguchi:20}, JVS, \cite{Tamaru:20} and
Tohoku \cite{Ogawa:21}  datasets, as  well as an  internal datasets composed  of 2  pop, 6
classical  singers.  This  dataset  contains  about 27h  of  singing  recordings from  136
singers. The last  dataset will be  denoted as \emph{SI}.

All  database  recordings  were  resampled  to 24kHz.   All  voice  files  were  annotated
automatically with F0 contours using the  FCN estimator \cite{Ardaillon:19a}. We emploi the
noise separation  algorithm described  in \cite{Huber:15a}  to separate  deterministic and
noise components and  calculate the noise/total energy  balance over 4 periods  of the F0.
We annotate segments with more than 50\% of the energy in the noise component as unvoiced.

For  the  optimization   we  use  Adam  optimizer  \cite{Kingma:15}   with  learning  rate
$lr=1e-4.$.  The  decay rate  parameters  are  $\beta_1=0.9, beta_2=0.999$,  for  training
without discriminator and $\beta_1=0.5, beta_2=0.5$ for training with discriminator. Batch
size is always 40, and the segment length is appoximately 200ms.

As objective functions we use the following loss functions.
The first loss is the f0 prediction loss given by 
\begin{equation}
  L_{F0} =(\sum_{k\in K} \|F_k-\hat{F}_k \|)/ \sum_k 1.
  \label{eqn:f0_pred_loss}
\end{equation}
$F_k$ is  the target F0 and  $\hat{F}_k$ are the  predicted value at time  sample position
$k\in K$ and $K$ is  the set of points that are annotated as  voiced and further than 50ms
away  from a  voiced/unvoiced boundary.  For these  unambiguously voiced  sections the  F0
predictor can be optimized using only the prediction error.

The second loss is a multi resolution spectral reconstruction loss similar to
 \cite{Jang:21}. It is composed of two terms the first one calculated as normalized
linear magnitude differences and the second as log amplitude differences.
\begin{eqnarray}
  L_{A} &=& \| |S| - |\hat{S}| \|_2/\| |S| \|_2, \mbox{and}
  \label{eqn:mr_linS_loss}            \\ 
  L_{L} &=& \frac{1}{KM} \| \log(S) - \log(\hat{S}) \|_1.
  \label{eqn:mr_logS_loss}       
\end{eqnarray}
Here $S$ and $\hat{S}$ are the  magnitudes of the STFT of the   target and generated
signals and $K$ and $M$ are the number of frames and the number of bins in the STFT matrices.

The final reconstruction loss is then the mean of the reconstruction losses obtained for
the different resolutions
\begin{equation}
  L_{R} = (\sum_{j} ( L_{A,j} +L_{L,j} ))/  \sum_j 1.
  \label{eqn:mr_recS_loss}             
\end{equation}
where $j$ runs over the resolutions. For the folloiwn gexperiements we used STFT with
window size in seconds given by $M\in[.02, 0.0375, 0.075]$, and hop size in seconds
$H\in[0.00375, 0.0075, 0.015]$.

The reconstruction loss  is used as objective  function for the pulse  shaping WaveNet and
for the  F0 predictor  around voiced  unvoiced boundaries (more  precisely within  50ms of
these boundaries  within voiced segments, and  within 20ms of these boundaries  in unvoiced
segements). In these transition areas we expect the F0 annotation to be less reliable and
not sufficient to create optimal resynhesis performance. Therefore here we optimize the
F0 predictor as part of the generator.

Finally,  when  training with  the  discriminator  loss $\L_D$  we  use  exactly the  same
discriminator configuration  and loss as  \cite{Jang:21}. The  only difference is  that we
only use 2 discriminator  rates. The first one working on the  original samplerate and the
second after  average pooling  of factor  4.  We motivate  the decision  to drop  the last
discriminator with  the fact that  the stability of  the periodic oscillations  is already
ensured  by the  excitation  model and  therefore  the discrinminator  is  only needed  to
evaluate  the pulse  form  and the  balance between  deterministic  and stochastic  signal
components.

For each model we first pretrain the F0  prediction model over 100k batches using only the
\glnref{eqn:f0_pred_loss}  as objective  function. Pretraining  the F0  predictor reliably
achieves prediction errors  below 3Hz. We wont  further discuss these.  As a  next step we
pretrain the full generator strtaing with  the pretrained F0 model loaded.  Pretraining of
the generator runs for 200k batches. To  create a generator without using adversarial loss
we  continue training  the  generator  for further  400k  iterations.  When training  with
adversarial loss we load the discriminator after the 200k training steps for the generator
and train with discriminator for further 800k batches.

\vspace{-2mm}
\subsection{Perceptual Evaluation}
\label{sec:evaluation}

In the following section will compare different models and configurations.  We will denote
these using a code  structured like: TXC. Here T will be replaced  by the model type using
the  following  shortcuts:  \textbf{MW}:  multi  band  excited  WaveNet  introduced  here,
\textbf{MMG}: multi band melgan from \cite{Yang:21}, \textbf{UMG} universal melgan vocoder
from \cite{Jang:21}. The further codes will be used only for the \textbf{MW} models, where
\emph{X} is either \textbf{U} for multi  voice (universal) models or \textbf{S} for single
voice models. Finally  \emph{C} is a sequence of letters  representing specific components
missing from  the full model. Here  only a single letter  is used. The letter  in the last
position  \emph{v} indicates  that the  model  does not  use  a dedicated  module for  VTF
prediction.  For the \textbf{MW} model we have trained
two  multi voice  models.  A  singing model  that  is trained  with the  full pitch  range
available in the  \emph{SI} dataset and a  speech model on the \emph{SP}  dataset.  In the
tables below  when singing data  is treated the singing  model is used.   Equivalently the
speech model will be used for speech.  During the perceptual evaluation we used the speech
mdoel as well for those singing signals that do stay in a pitch range thet was part of our
speech databases. For that special cases we denote the model as \textbf{MWU$_{SP}$}.

We first  summarize results  about pretraining. Pretraining  the F0 models  on any  of the
datasets converges reliably  to an F0 prediction  error around 2Hz for  singing and around
2.5Hz for speech. Pretraining the generator achieves spectral reconstruction errors in the
order of 3dB for singing and 3.2dB for speech. Reconstruction error on the mel spectrogram
is even smaller and generally $<$2dB. Listening to the generated sounds reveals a constant
buzz in many of the noise sections. The main problem here are residual pulses that are not
sufficiently suppressed by  the pulse forming wavenet.  To solve this problem  we will use
the time domain discriminators proposed originally in \cite{Kumar:19}.

\vspace{-3mm}
\subsection{Perceptual tests}

We have conducted a  perceptual test evaluating the perceived quality  of the selected MWU
and MWS  models trained on multi  and single user database.   We will use seen  and unseen
speakers, languages,  expressivities, as well as  singing styles.  For these  tests we use
three     baselines.      We     used     an     open     source     multi-band     melgan
implementation\footnote{\url{https://github.com/TensorSpeech/TensorflowTTS}} and  trained it for
1M iterations  on the \emph{DI}  and \emph{LJ}  datasets.  Further we  downloaded original
samples   together   with   resynthesized   results  of   the   Universal   MelGan   model
\cite{Jang:21}\footnote{\url{https://kallavinka8045.github.io/icassp2021/}} and used  these as a
baseline  for the  multi  voice  models.  Each  of  the tests  has  been  conducted by  42
participants, consisting  of audio and  music professionals working  at or with  IRCAM and
native English speakers recruited via the prolific online platform\footnote{Demo sounds are
available under \url{http://recherche.ircam.fr/anasyn/roebel/MBExWN_demo/}.}.

In  contrast to  perceptual tests  performed in  other studies  our main  interest is  the
perceptually transparent resynthesis of the original speech signal.  Therefore we chose to
perform a MUSHRA test containing the reference  signal and a group of resynthesized signal
that the  participants can play  as they like.  The task given  was to concentrate  on any
differences that might be  perceived between the original and the  resynthesis and to rate
the peceived differences  on a scale from  0 to 100 with categories imperceptible
(80-100),  perceptible, not  annoying  (60-80),  slightly annoying  (40-60),  annoying
(20-40), very annoying  (0-20).  Results are listed in  table \tabref{tab:perc_test}.  The
first column indicates the  data source the data is taken from.   The second column marked
\textbf{HREF} represents a hidden reference (copy of the reference) for which we
expect and observe an evaluation  around 90 for all cases.   In the second column we  find the MBExWN
models  trained on  a multi  voice dataset.  In the  sub sequent  columns we  find various
baselines.

In the upper part of the table we  find the perceptual evaluation of singing data.  In the
first line the evaluation data of the \emph{SI}  dataset is used. The result of the MWExWN
model trained  on the  singing voice  dataset is equivalent  to the  result of  the hidden
reference.  In  the second line  we compare the multi  singer model with  dedicated single
singer models trained on  the \emph{DI} dataset as well as the  multi speaker model.  Note
that the singer \emph{DI} is neither part \emph{SI} nor of \emph{SP}. The best result here
is obtained by the  the single singer MBExWN model.  This model  is trained exclusively on
that singer.   Both multi  voice models, whether  trained on sining  or trained  on speech
achieve a  rating in the highest  category of the  mushra test. The model  achieving worst
performance is the  multi-band melgan trained as well on  \emph{DI}.  The explanation here
are  instabilities in  the synthesized  F0 trajectory,  sometimes the  model even  changes
phonemes.  This  problem confirms our observation  with the existing neural  vocoders that
tend to have probems with stable periodic oscillations.   In the last two lines we use pop
and metall solo singing as out of domain singing. Both multi voice models degrade somewhat
for pop singing  and produce an annoying  quality for saturated metal  singing.  The model
structure will  need to change  to support the  characteristic sub harmonics  of saturated
voices. In the lower part of the table we  find results for speech data. In the first line
we see the multi speaker model evaluated on  its validation data. The model remains in the
range of the highest MUSHRA quality. A  likely reason for the sligh reduction are frequent
pop  noises in  the  speech signals.   The  model will  not  reproduce these  perceptually
transparent. In the  second line we find the  evaluation of the multi speaker  model on an
unseen  speaker comparing  to the  MBExWN and  MMG single  speaker models  trained on  the
evaluated speaker the  MWS model is rated  best but the differences  are not statistically
significant.  The last model in the test is a MBExWN multi speaker model configure without
the VTF component.   This model is clearly perceived as  less transparent, which validates
the VTF as  part of the model structure.   In the last line we compare  the data retrieved
from the UMG demo  site.  The samples are considered out  of domain samples (expressivity,
language, speakers) for both the UMG and the  MWU model. Comparing these two shows that the
MWU model is ranked higher as the UMG model but this difference may not be significant.

\begin{table}[ht!]

 \centering
\begin{tabular}{l@{\hspace{3mm}}c@{\hspace{3mm}}c@{\hspace{3mm}}c@{\hspace{3mm}}c@{\hspace{3mm}}c@{\hspace{1mm}}}
  \toprule  
\multicolumn{6}{l}{Singing Models/Singing Data}\\ \toprule
  Data   & HREF      &  MWU           & MWS & MMG       & MWU$_{SP}$ \\ \midrule
  SI     & 90 (4.2)  & 90 (3.8)         &   -    &   -    &  - \\ 
  DI     & 89 (4.6)  & \emph{83 (5.8)}  & \textbf{89 (3.8)} & 66 (8.4) &  \emph{80 (6.5)} \\   \midrule

  Pop-sing & 88 (4.7)  & \emph{71 (8.4)} & -  &   - & \textbf{\emph{76 (7.8)}} \\
  Met-sing & 91 (2.8)  & \textbf{\emph{57 (9.9)}} & - &   -  & \emph{55 (8.8)} \\
  \bottomrule
  \toprule  
\multicolumn{6}{l}{Speech Data/Speech Models}\\ \midrule
  Data   & HREF          &  MWU           & MWS             & MMG      &  MWUv   \\ \midrule
  SP     & 92 (2.8)      & 85 (7.3)         &  -     I           &  -       &  -     \\
  LJ     & 90 (3.6)      & \emph{84 (6.1)}  & \textbf{85 (5.1)} & 83 (5.6) &  77 (6.9) \\ \toprule
  Data   & HREF          & MWU            &    UMG   \\ \midrule
  UMG\_V & 92 (5.1)      & \textbf{\emph{84 (6.3)}}  & \emph{79 (8.7)} \\
  \bottomrule
\end{tabular}

\caption{Perceptual evaluation of the perceived difference beween original and resynthesis
  for different models and conditions.}
\label{tab:perc_test}
\end{table}

\subsection{Complexity}

The multi speaker model with 320 WaveNet channels  has  about 10M parameters and
achieves inference speed of 50kSamples/s when running on a single
core of an  Intel i7 laptop CPU. On a NVidia
V100 GPU the inference rate of 2.4Msamples/s. These numbers
compare favourably  with the universal melgan  \cite{Jang:21} that has 90M  parameters  and
achieves an inference speed of 860kHz on a V100 NVidia GPU.

\section{Conclusions}

In this  paper we have presented  MBExWN a new  neural vocoder with an  externally excited
WaveNet as source. A  perceptual test has shown that the  proposed model supports achieves
near tranparent  quality even  for out  of domain  data. The  signal degrades  when
confronted with rough and saturated voices. Further research will be conducted to
solve these cases.

\subsection{Acknowledgements}

We would like to thank Won Jang for sharing information and materials related to the
universal melgan.

\vfill\pagebreak

\bibliographystyle{IEEEbib}
\bibliography{allrefs}

%\listoftodos

\end{document}